\pdfoutput=1

\documentclass[
    ,final            
  ]
  {aipproc}

\layoutstyle{8x11double}

\usepackage{amsmath}
\usepackage{amssymb}
\usepackage{amsthm}

\newcommand{\numax}{\mbox{$\nu_\mathrm{max}$}}

\begin{document}


\title{Asteroseismology of red-clump stars with CoRoT\footnote{The CoRoT space mission was developed and is operated by the French space agency CNES, with participation of ESA's RSSD and Science Programmes, Austria, Belgium, Brazil, Germany, and Spain.}\ and Kepler}



\author{A. Miglio}{
  address={Institut d'Astrophysique et de G\'eophysique de l'Universit\'e de Li\`ege,
All\'ee du 6 Ao\^ut, 17 B-4000 Li\`ege, Belgium}
,altaddress={Postdoctoral Researcher, Fonds de la Recherche Scientifique - FNRS, Belgium} 
}

\author{J. Montalb{\'a}n}{
  address={Institut d'Astrophysique et de G\'eophysique de l'Universit\'e de Li\`ege,
All\'ee du 6 Ao\^ut, 17 B-4000 Li\`ege, Belgium}}

\author{P. Eggenberger}{
  address={Observatoire de Gen\`eve, Universit\'e de Gen\`eve, 51 chemin des Maillettes, 1290 Sauverny, Switzerland}}

\author{S. Hekker}{
  address={School of Physics and Astronomy, University of Birmingham, Edgbaston, Birmingham B15 2TT United Kingdom}
  ,altaddress={Instituut voor Sterrenkunde, K.U. Leuven, Celestijnenlaan 200D, 3001 Leuven, Belgium}} 


\author{A. Noels}{
  address={Institut d'Astrophysique et de G\'eophysique de l'Universit\'e de Li\`ege,
All\'ee du 6 Ao\^ut, 17 B-4000 Li\`ege, Belgium}}

\begin{abstract}
The availability of asteroseismic constraints for a large number of red giants with CoRoT and, in the near future with Kepler, paves the way for detailed studies of populations of galactic-disk red giants.
We investigate which information on the observed population can be recovered by the distribution of the observed seismic constraints: the frequency of maximum power of solar-like oscillations (\numax) and the large frequency separation ($\Delta\nu$).  We use the distribution of \numax\ and of $\Delta\nu$ observed by CoRoT in nearly 800 red giants in the first long observational run, as a tool to investigate the properties of galactic red-giant stars through the comparison with simulated distributions based on synthetic stellar populations.
   We can clearly identify the bulk of the red giants observed by CoRoT as red-clump stars, i.e. post-flash core-He-burning stars. The distribution of \numax\ and of $\Delta\nu$ give us access to the distribution of the stellar radius and mass, and thus represent a most promising probe of the age and star formation rate of the disk, and of the mass-loss rate during the red-giant branch.
    This approach will be of great utility also in the interpretation of forthcoming surveys of
    variability of red giants with CoRoT and Kepler. In particular, an asteroseismic mass estimate of clump stars in the old-open clusters observed by Kepler, would represent a most valuable observational test of the poorly known mass-loss rate on the giant branch, and of its dependence on metallicity.



\end{abstract}

\maketitle


\section{Introduction}
The clump of red giants in the color-magnitude diagram (CMD) of intermediate-age and old open clusters was recognized as being formed by stars in the stage of central helium burning since the works of \citet{Cannon70} and \citet{Faulkner73}. The near constancy of the clump absolute magnitude in these clusters was correctly interpreted as the result of He ignition in an electron-degenerate core. Under these conditions, He burning cannot start until the stellar core mass attains a critical value of about 0.45 $\rm M_\odot$. From this, it follows that low-mass stars developing a degenerate He core after H exhaustion, have similar core masses at the beginning of He burning, and hence similar luminosities.

Beside old-open clusters, the red giant clump (RC) is also a remarkable and well populated feature in the CMD of composite stellar populations, like galaxies. A major advance in our understanding of RC stars was possible thanks to the determination of precise parallaxes with the \emph{Hipparcos} mission, which allowed us to identify clearly the clump in the CMD of nearby stars \citep{Perryman97}.

Thanks to its near-constant luminosity, the RC gained importance also as a distance indicator providing distance estimates to the galactic centre \citep{Paczynski98} and nearby galaxies such as Fornax \citep{Rizzi07}, M33 \citep{Kim02} and LMC \citep[see e.g.][]{Girardi01, Salaris03}.

In the above-mentioned examples red clump stars were identified in simple or composite stellar populations of a given distance.
With the asteroseismic constrains that are now becoming available, we are able to characterize the population in the galactic disk even without any
information about distance. In the first section, we show how very simple seismic
observables, such as the distribution of \numax, can be used to
identify the population of red giants observed by CoRoT as being
dominated by the red clump of the galactic disk. This approach could also be
applied to the numerous red giants that will be monitored with the Kepler satellite and, as we show in the second section, could be of major relevance in the study of stellar evolution of simpler populations, such as those of the old-open clusters, that Kepler will observe.



\section{The population of galactic-disk red giants seen with CoRoT eyes}
The CoRoT satellite \citep{Baglin06a} is providing high-precision, long and uninterrupted photometric monitoring of thousands of stars in the fields primarily dedicated to the search for exoplanetary systems (EXOField). These observations reveal a wealth of information on the  properties of solar-like oscillations in a large number of red giants \citep[see][]{DeRidder09}, i.e. their oscillation frequency range, amplitudes, lifetimes and nature of the modes (radial as well as non-radial modes were detected). The goldmine of information represented by this detection is currently being exploited.

As described in detail by \citet{Hekker09}, about 800 solar--like pulsating red giants were identified in the field of the first CoRoT 150-days long run in the direction of the galactic centre (LRc01).
\citet{Hekker09} searched for the signature of a power excess due to solar-like oscillations in all stars brighter than $m_{\rm V}$=15 in a frequency domain up to 120 $\mu$Hz. They found that the frequency corresponding to the maximum oscillation power (\numax) and the large frequency separation of stellar pressure modes ($\Delta\nu$) are non-uniformly distributed.

The distributions of \numax\ and $\Delta\nu$ show a single dominant peak located at $\sim30$ $\mu$Hz and $\sim4$ $\mu$Hz, respectively (the corresponding histograms are also shown in the lower panels of Fig. \ref{fig:numax} and \ref{fig:dnu}). We note, however, that the detection of solar-like pulsations in the low-frequency end ($\nu_{\rm max} \lesssim 20$ $\mu$Hz) is compromised due to the long-period trends, the activity and granulation signal. We should therefore regard the observed distribution at the lowest frequencies as strongly biased \citep[see][]{Hekker09}.

We focus on the simplest and most robust seismic constraints provided by the observations: \numax\ and $\Delta\nu$, and on the information they supply on the stellar parameters.
As suggested by \citet{Brown91}, \numax\ is expected to scale as the acoustic cutoff frequency of the star, which defines an upper limit to the frequency of acoustic oscillation modes. Following \citet{Kjeldsen95} and rescaling to solar values, \numax\ can be expressed as :
\begin{equation}
\nu_{\rm max}=\frac{M/M_\odot}{(R/R_\odot)^2\sqrt{T_{\rm eff}/5777 K}}\ \rm3050\ \mu Hz.
\label{eq:numax}
\end{equation}

The large frequency separation of stellar p modes, on the other hand, is well known to be inversely proportional to the sound travel time in the stellar interior, and is therefore related to the mean density of the star by the following expression \citep[see e.g.][]{Kjeldsen95}:

\begin{equation}
\Delta\nu=\sqrt{\frac{M/M_\odot}{(R/R_\odot)^3}}\ 134.9\ \rm \mu Hz.
\label{eq:dnu}
\end{equation}

As presented in \citep{Miglio09} (Paper I), we investigated the properties of the underlying stellar population that determine the observed \numax\ and $\Delta\nu$ distributions, and whether the presence of a single peak in \numax\ and $\Delta\nu$ is compatible with current models of red-giant populations in the galactic disk.

\subsection{Stellar population synthesis}
\label{sec:pop}
In order to simulate the composite stellar population observed in the LRc01 of CoRoT's EXOField we used the stellar population synthesis code {\sc trilegal} \citep{Girardi05} designed to simulate photometric surveys.

\begin{figure}
  \includegraphics[width=\columnwidth]{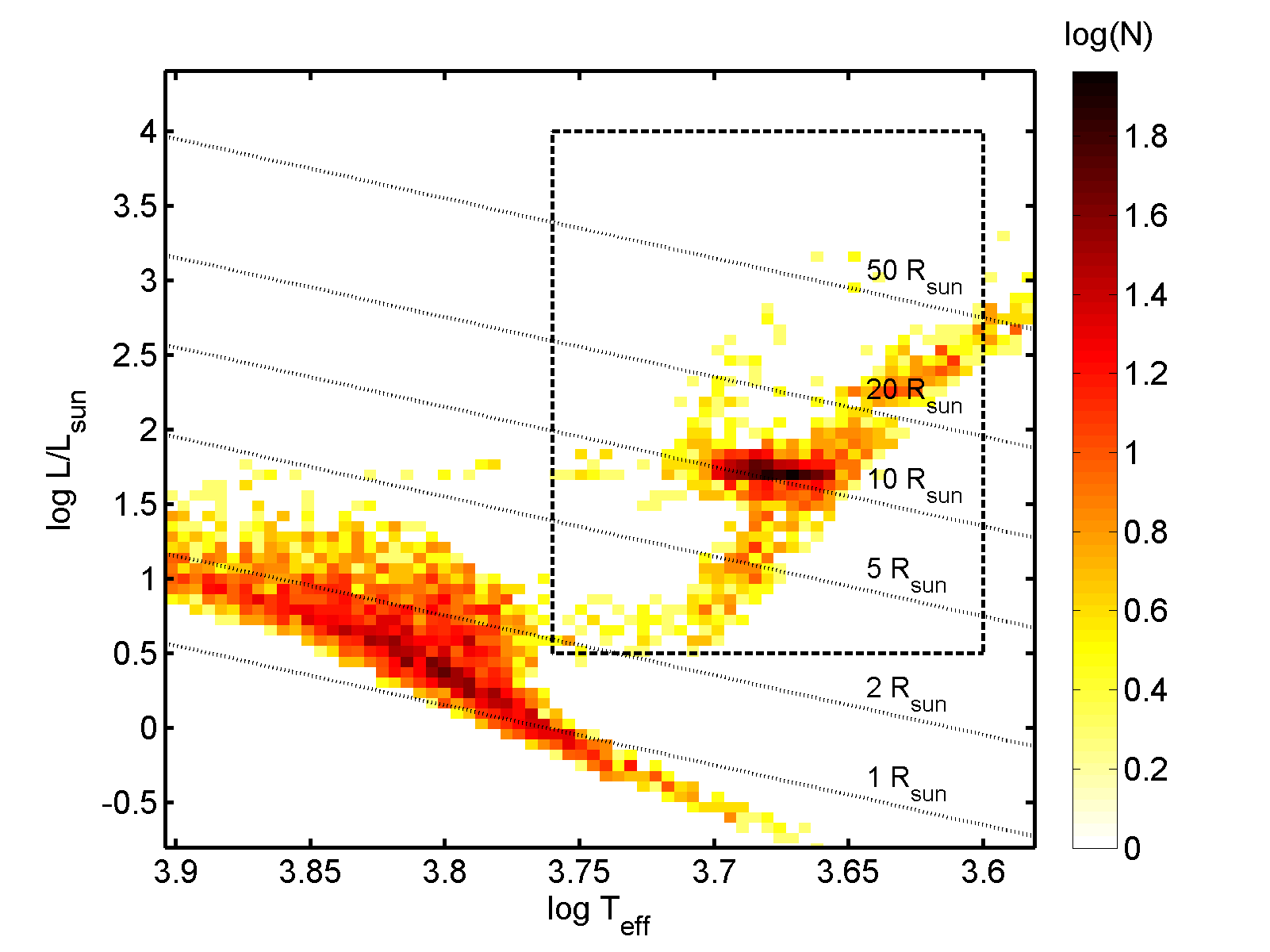}
      \caption{Theoretical Hess diagram of the population T1 simulated with {\sc trilegal}. The dashed rectangle delimits the location in the HR diagram of the population of G-K giants considered in this work. The bar on the right shows, in a color-coded logarithmic scale, the number of stars per $\log{L/L_\odot}-\log{T_{\rm eff}}$ bin.}
         \label{fig:hess}
\end{figure}

In {\sc trilegal} several model parameters (such as the star formation history and the morphology of different galactic components) were calibrated to fit Hipparcos data \citep{Perryman97}, as well as star counts from deeper photometric surveys.

In our {\sc trilegal} simulation (T1) we adopted the standard parameters for the four components of the Galaxy (halo, bulge, thin and thick disk). We simulated the stellar population in the sky area observed by CoRoT during LRc01 (3.9 deg$^2$ centered at $l=37^\circ$, $b=-6.9^\circ$) and considered stars in the observed magnitude range ($11 \lesssim m_{\rm V} \lesssim 15$). This population turns out to be dominated by thin-disk stars.

The default star formation rate in the thin disk is assumed to be constant over the last 9 Gyr, and the age-metallicity relation is taken from \citet{Rocha00}. An exhaustive description of the evolutionary tracks used in the simulations is given by \citet{Girardi05} and references therein. 

An HR diagram showing the number density of stars (Hess diagram) of the synthetic population is presented in Fig. \ref{fig:hess}, where the main sequence and the red clump (RC) clearly appear as the most populated areas.

We shall henceforth restrict our interest to the population of G-K giants delimited by the dashed rectangle in Fig. \ref{fig:hess}. In the HR diagram (see Fig. \ref{fig:hess}) the distribution of giants is not uniform and predominantly peaked in the red clump, near iso-radii lines corresponding to $R \simeq 10$ $R_\odot$.

\subsection{Observed vs. simulated \numax\ and $\Delta\nu$ distributions}
\label{sec:results}
We now simply evaluate, by means of Eq. \ref{eq:numax} and \ref{eq:dnu}, the theoretical \numax\ and $\Delta\nu$ distributions based on the properties ($R$, $M$, $T_{\rm eff}$) of the synthetic population presented in the previous section.
Among all the stars in the population we select G-K giants ($0.5 < \log{L/L_\odot} < 4$ and $3.6 < \log{T_{\rm eff}} < 3.76$, see Fig. \ref{fig:hess}). As we previously discussed, we should restrict the comparison to \numax\ $\gtrsim 20$ $\mu$Hz.

\begin{figure}
\begin{minipage}{\columnwidth}
\centering
\includegraphics[width=\columnwidth]{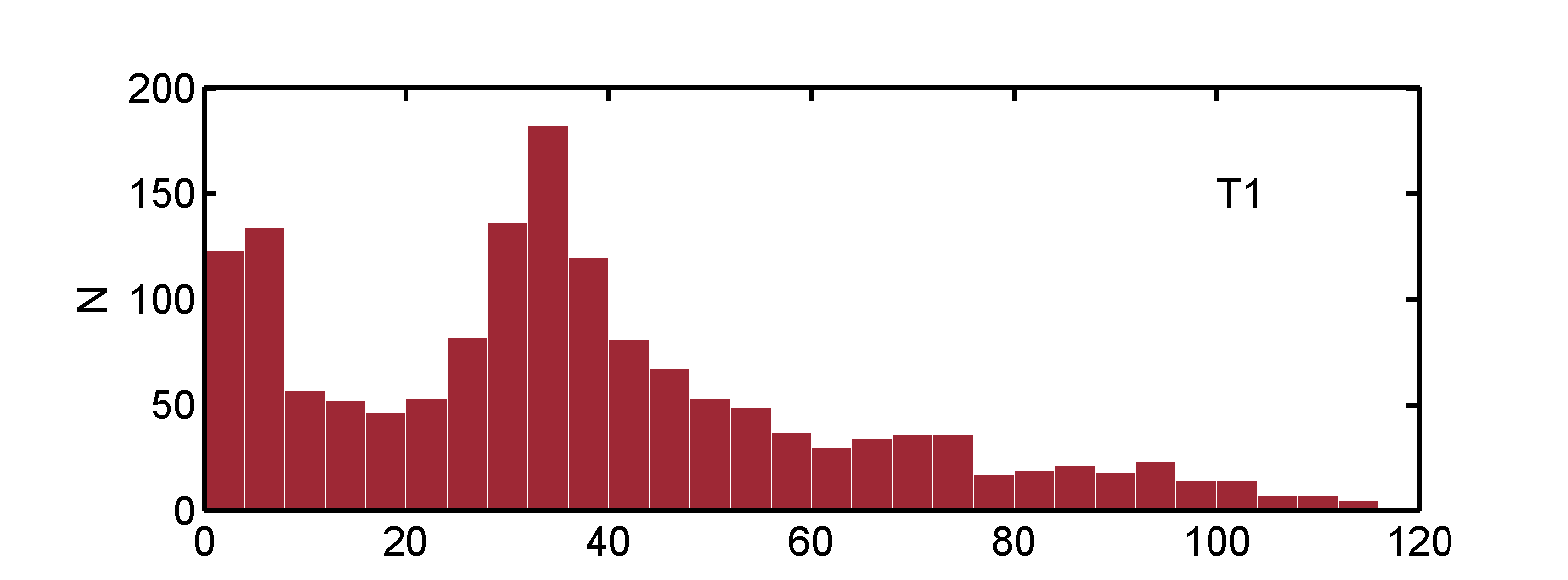}
\includegraphics[width=\columnwidth]{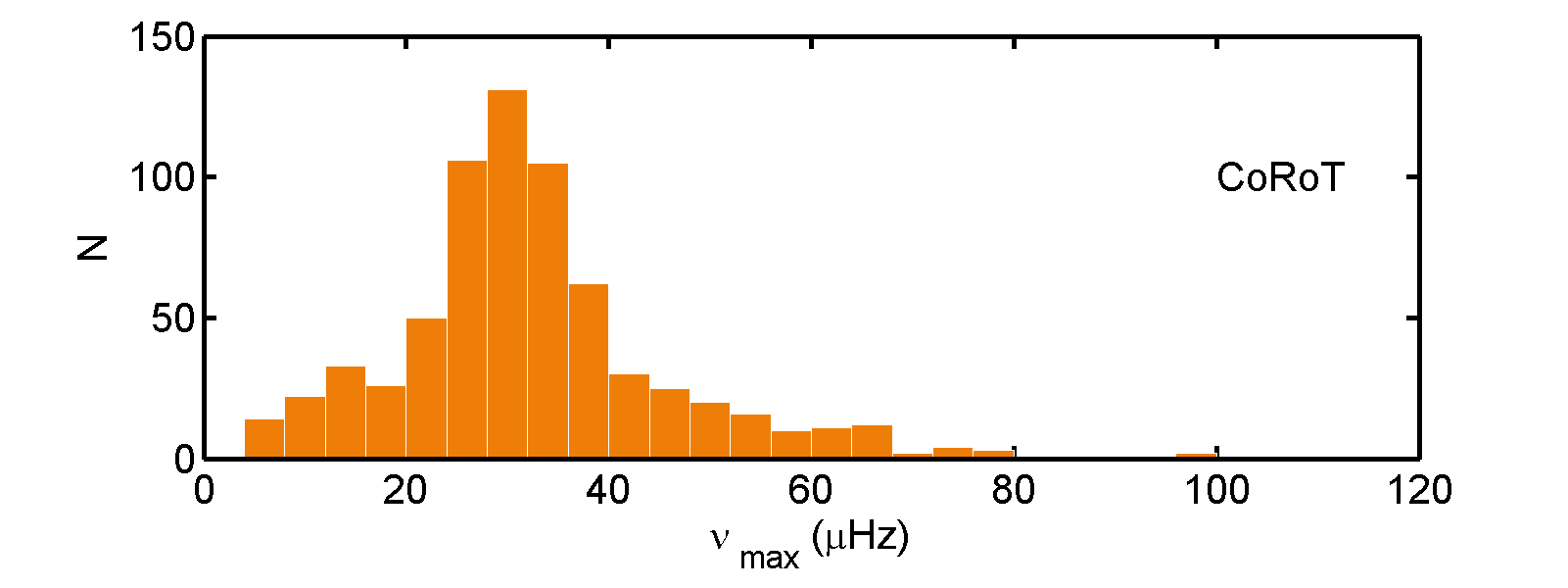}
\end{minipage}
   \caption{Histogram showing the comparison between the \numax\ distribution of the observed (\emph{lower panel}) and simulated populations of red giants (\emph{upper panel}).
              }
         \label{fig:numax}

\end{figure}

The \numax\ distribution we obtain for T1 is reported in the upper panel of Fig. \ref{fig:numax}. It is clear that the dominant maximum in the observed histogram (lower panel of Fig. \ref{fig:numax}) can be qualitatively well reproduced by the stellar population model.

\begin{figure}
  \includegraphics[width=\columnwidth]{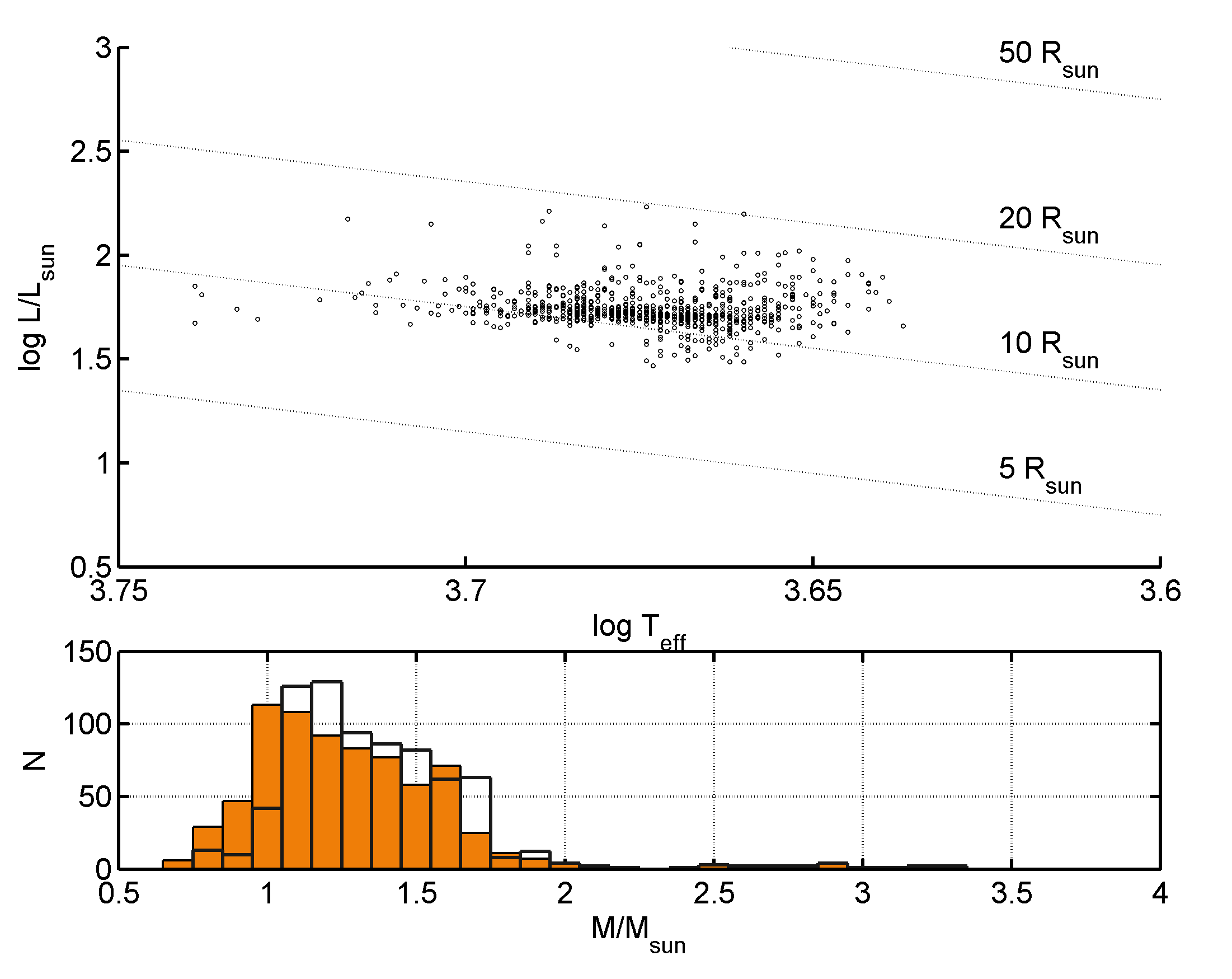}
   \caption{\emph{Upper panel:} HR diagram showing the position of stars in the population T1 with 20 $\mu$Hz $< \nu_{\rm max} < 50$ $\mu$Hz. The radius distribution of these stars has a mean of 11.2 $R_\odot$ with a standard deviation of 1.5 $R_\odot$.
           \emph{Lower panel:} The mass distribution of the sub-population in the upper panel is represented by full bars, whereas empty bars illustrate the mass distribution of their progenitors.}
         \label{fig:numax2060}
\end{figure}

The comparison with the observed \numax\ distribution  suggests that the bulk of CoRoT LRc01 pulsating giants are red clump stars. In Fig. \ref{fig:numax2060}, upper panel, we show the location in the HR diagram of stars belonging to T1 and having 20 $\mu$Hz $< \nu_{\rm max} < 50$ $\mu$Hz: this lets us identify the dominant peak in the distribution as due to red-clump stars, i.e. low-mass stars with actual masses $0.8 \lesssim M/M_\odot \lesssim 1.8$ (see the lower panel of Fig. \ref{fig:numax2060}), that are in the core He-burning phase after having developed an electron-degenerate core during their ascent on the RGB and passed the He-flash.

\begin{figure}
\begin{minipage}{\columnwidth}
\centering
\includegraphics[width=\columnwidth]{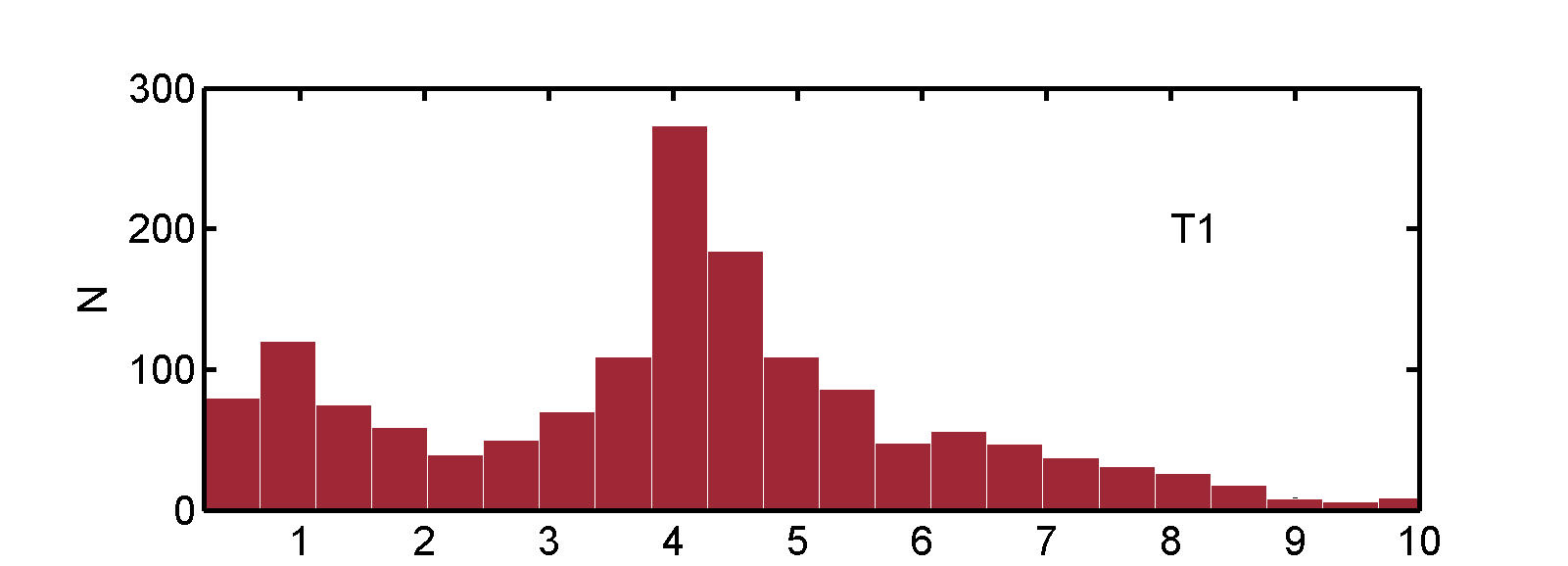}
\includegraphics[width=\columnwidth]{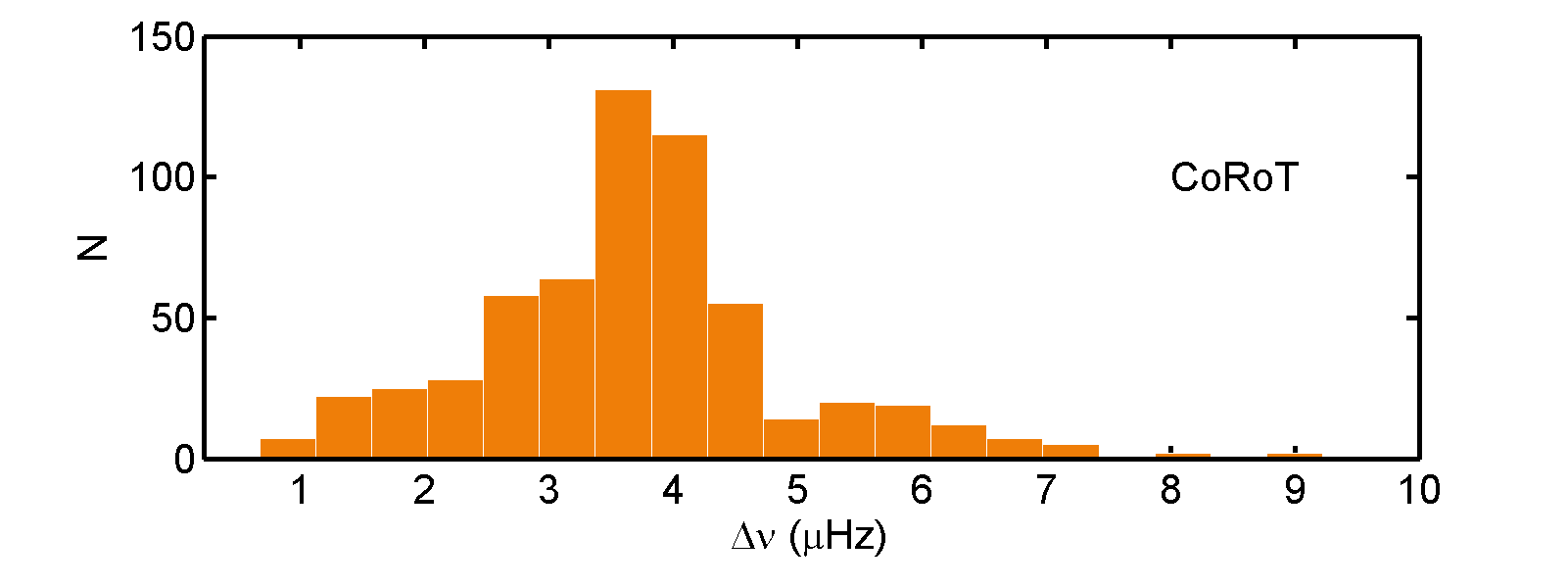}
   \caption{Histogram showing the comparison between the $\Delta\nu$ distribution of the observed (\emph{lower panel}) and simulated populations of red giants (\emph{upper panel}).
              }
         \label{fig:dnu}
\end{minipage}
\end{figure}

This conclusion is also fully supported by the comparison between the observed and theoretical distribution of $\Delta\nu$, as shown in Fig. \ref{fig:dnu}.

The observed dominant peaks in the \numax\ and $\Delta\nu$ distributions are qualitatively explained by the models. Nevertheless, a Kolmogorov-Smirnov test shows that the discrepancies between the simulated and observed distributions are, from a statistical point of view, highly significant due to the fact that the samples contain a large number of stars.

The predicted \numax\ distribution is affected by several parameters in stellar populations models.
We note however that, since  most of the stars belonging to the RC cover a limited range in radius, \numax\ represents a valuable tracer of the mass distribution of the population (see Fig. \ref{fig:numax2060}).
The mass distribution of RC stars is, in its turn, highly dependent on the age and, more generally, on the star formation rate of the composite population, as well as on the mass-loss rate adopted during the RGB (see Paper I for a more detailed discussion).

\section{Future observations with CoRoT and Kepler}
\label{sec:future}
Asteroseismic constraints for populations of red giants will soon be available for various fields in the Galaxy.

CoRoT is observing in G-K giants in several fields near the galactic equator, with longitudes near the galactic centre ($l\simeq 35^\circ$) and anticenter ($l\simeq 215^\circ$). The Kepler satellite \citep{JCD08} will monitor red giants in a larger field located at $l \simeq 75^\circ$ and $7^\circ \lesssim l \lesssim 20^\circ$ (see \texttt{http://astro.phys.au.dk/KASC/}).

In addition to the observations of red giants belonging mostly to the composite stellar population of the galactic disk, Kepler will also perform observations of solar-like oscillations in simple, well constrained populations: the two old-open clusters NGC6791 and NGC6819.

\subsection{Red clumps in simple populations: NGC6791 and NGC6819}
NGC6791 and NGC6819 are well studied old-open clusters, both showing a red clump in their color-magnitude diagrams, as reported e.g. in \citet{Grocholski02} and \citet{Kalirai04}.

To have a first estimate of the expected seismic properties of the red giants in the two clusters, we computed a preliminary simulation by using the code {\sc basti} \citep{Pietrinferni04}, assuming single-age single-metallicity populations. In the case of NGC6791 we adopted as age 9 Gyr and  [Fe/H]=0.4 \citep[see][and references therein]{Grundahl08} and for NGC6819 an age of 2.5 Gyr and solar metallicity \citep{Kalirai04}.

\begin{figure}
\centering
\includegraphics[width=\columnwidth]{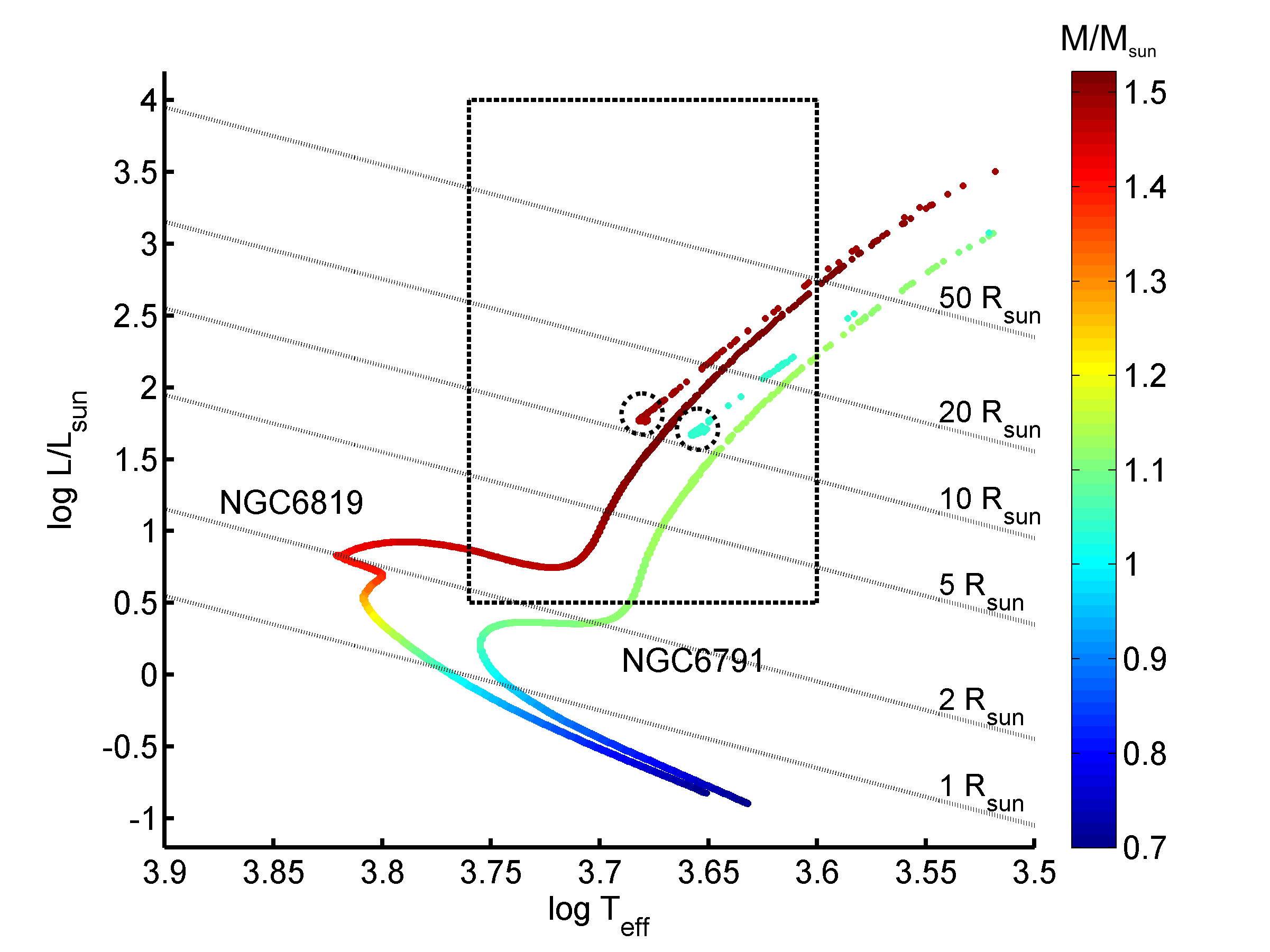}
\label{fig:hrclusters}
\caption{HR diagram of the one-age one-metallicity populations representing the clusters NGC6819 and NGC6791. The mass of the stars is color-coded (see color bar on the right). The dashed rectangle delimits the region of G-K giants we considered, while the dashed circles indicate the location of the red clump in the two clusters.}
\end{figure}

The position in an HR diagram of the stars of the synthetic populations of NGC6791 and NGC6819 is shown in Fig. \ref{fig:hrclusters}. The mass of the stars is color-coded in the plot, and indicates a mass of turnoff stars of about 1.1 and 1.4 $M_\odot$ for NGC6791 and NGC6819, respectively. The location of the red clump of the clusters is encircled by dashed lines.

\begin{figure}
\begin{minipage}{\columnwidth}
\centering
\includegraphics[width=.9\columnwidth]{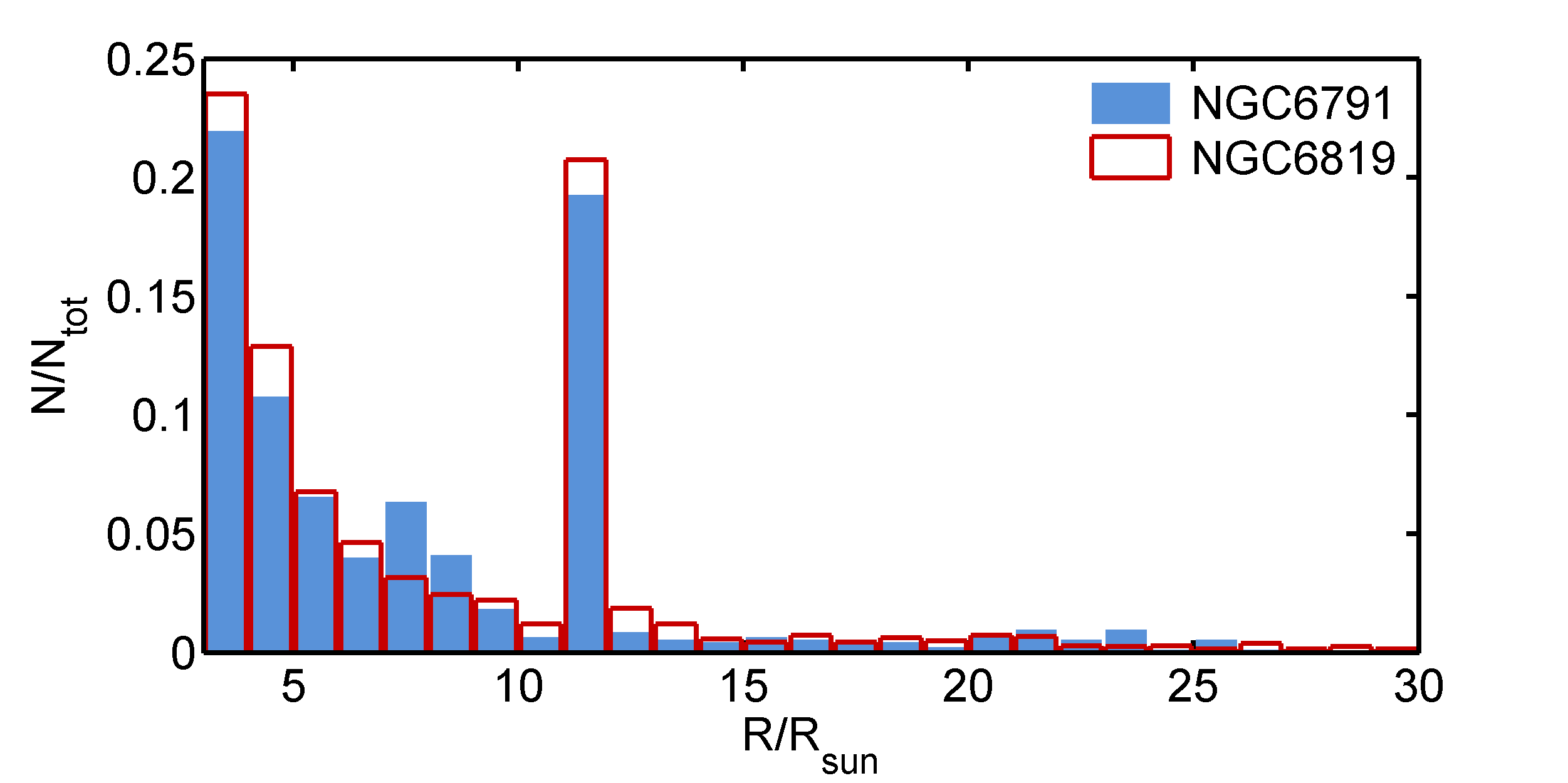}
\includegraphics[width=.9\columnwidth]{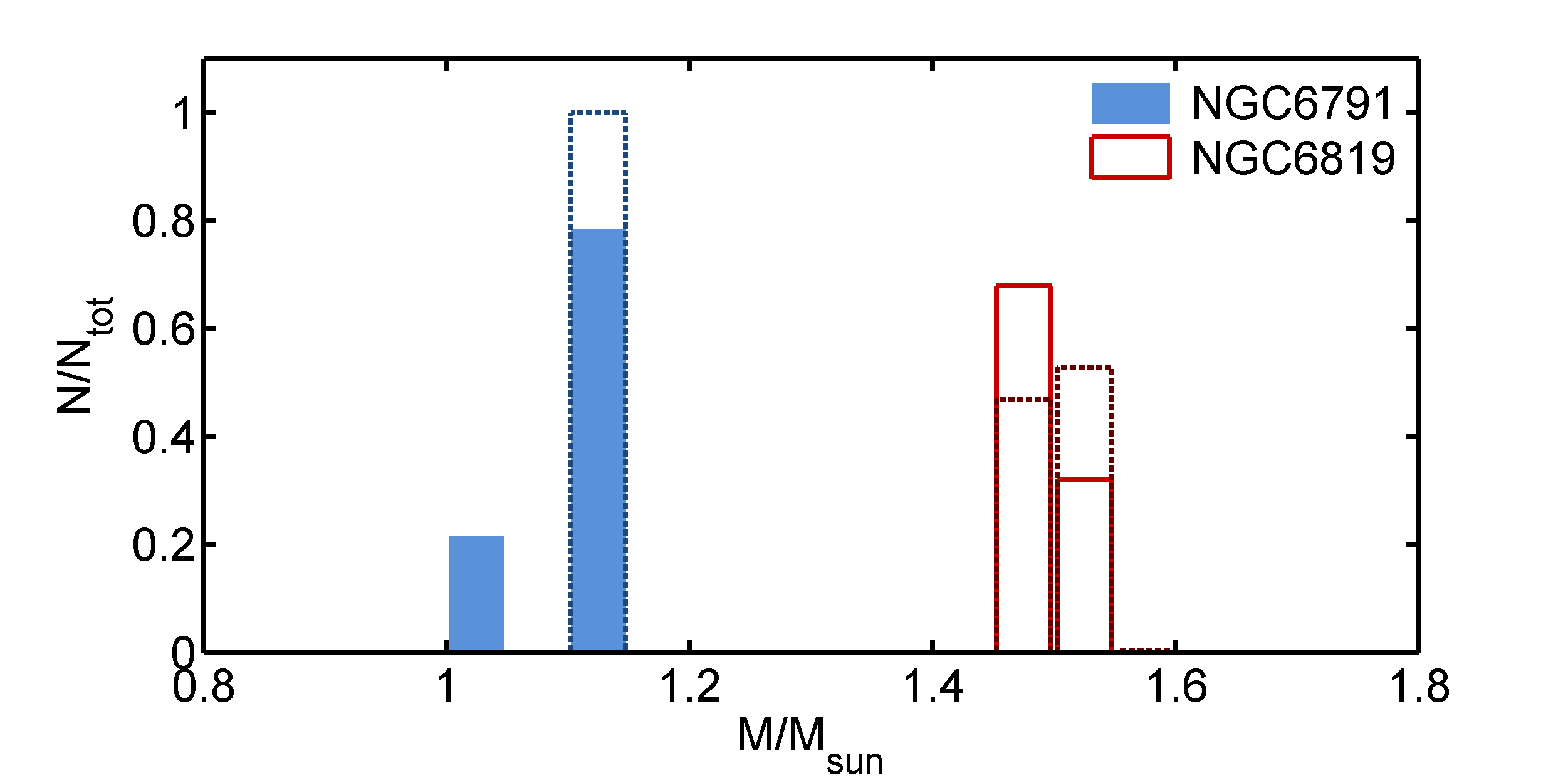}
\includegraphics[width=.9\columnwidth]{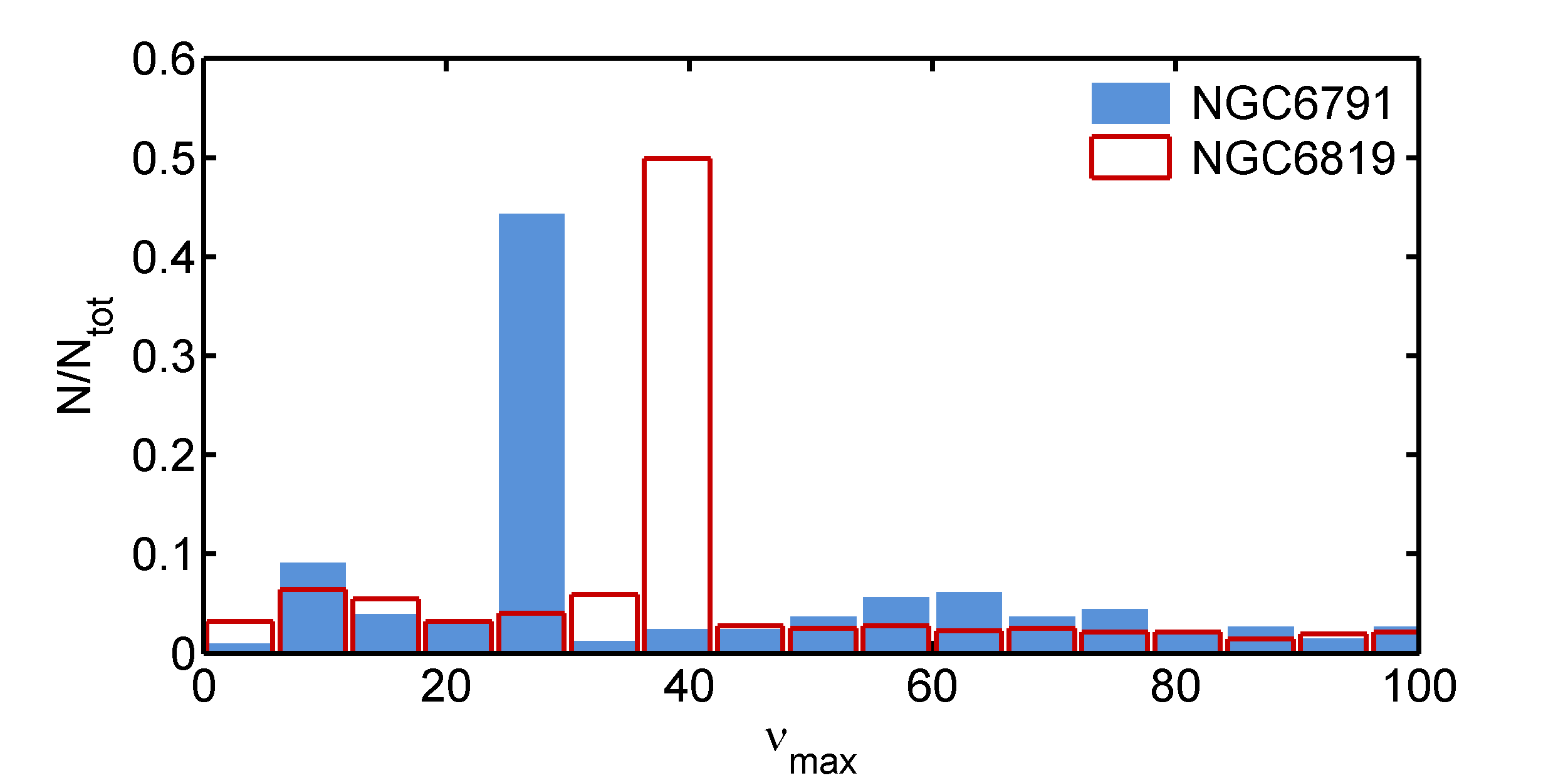}
\end{minipage}
\label{fig:bars}
\caption{{\it Upper panel:} Radius distribution of G-K giants in the synthetic population of NGC6791 and NGC6818. {\it Central panel:} full lines indicate the mass distribution of the stars, while dotted lines the mass distribution of their progenitors. {\it Lower panel:} \numax\ distribution for the red-giant populations in the two clusters with \numax\ $< 100$ $\mu$Hz. The significantly different \numax\ corresponding to the isolated peaks traces the mass difference of red-clump stars in NGC6791 and NGC6818 as predicted from the models.}
\end{figure}

In Fig. \ref{fig:bars} we present the radius, mass and \numax\ distributions of the red giants delimited by the dashed rectangle in Fig. \ref{fig:hrclusters}. As shown in Fig. \ref{fig:bars}, upper panel, the radius distribution of G-K giants is very similar in the two clusters, and is sharply peaked between 11 and 12 $R_\odot$, in correspondence to red-clump stars. As a consequence (see Eq. \ref{eq:numax}), the distribution of \numax\ is a reliable tracer of the mass distribution of red-clump stars. The latter is determined by the age of the population in the first place, but it is also affected by the mass-loss rate adopted during the red-giant-branch (see central panel of Fig. \ref{fig:bars}).

Since the age of the clusters can be determined by other means, e.g. by isochrone-fitting \citep[see][for a discussion]{Grundahl08}, it is clear that the asteroseismic mass estimate of clump stars represents a most valuable observational test of the poorly known mass-loss rate on the giant branch, and of its dependence on metallicity \citep[see e.g.][and references therein]{Carraro96}.

\section{Concluding remarks}
The detection of solar-like oscillations in populations of red giants with CoRoT allows us to access the properties of (so far) poorly-constrained populations in the galactic disk.

We obtain a robust identification of the bulk of CoRoT LRc01 red giants as red clump stars. This conclusion is based only on seismic constraints, namely on the comparison between the \numax\ and $\Delta\nu$ distributions observed and those based on a synthetic stellar population. The latter was computed with the population synthesis code {\sc trilegal}, assuming standard parameters \citep[see][]{Girardi05} to describe the morphology and the star formation history of the Galaxy components.

This approach will be of great utility in the interpretation of other surveys of
variability of red giants with CoRoT and Kepler. In  particular in the case of red giants in old-open clusters,
where the number of free parameters  affecting the theoretical distributions
of \numax\  and $\Delta\nu$  is significantly reduced. This should allow us
to constrain other important aspects of stellar evolution such as
the mass-loss  rate during the evolution in the red giant branch, or
transport processes during main-sequence evolution.


\begin{theacknowledgments}
  This work has made use of {\sc trilegal} and {\sc basti} web tools. A.M. is thankful to the Belgian FRS-FNRS for financial support.

\end{theacknowledgments}



\bibliographystyle{aipproc}   

\bibliography{../../andrea}

\IfFileExists{\jobname.bbl}{}
 {\typeout{}
  \typeout{******************************************}
  \typeout{** Please run "bibtex \jobname" to optain}
  \typeout{** the bibliography and then re-run LaTeX}
  \typeout{** twice to fix the references!}
  \typeout{******************************************}
  \typeout{}
 }

\end{document}